\documentclass[prl,twocolumn,showpacs,preprintnumbers,amsmath,amssymb]{revtex4}

\usepackage{graphicx}

\begin{document}

\title{Fighting cancer with virus}

\author{S. C. Ferreira Jr.}

\author{M. L. Martins}
\email{mmartins@mail.ufv.br}

\author{M. L. Vilela$^1$}

\affiliation{Departamento de F\'{\i}sica, Universidade Federal de Vi\c{c}osa, Vi\c{c}osa,
MG, Brazil\\
$^1$Departamento de Biologia Animal, Universidade Federal de Vi\c{c}osa, Vi\c{c}osa,
MG, Brazil}

\date{\today}

\begin{abstract}
One of the most promising strategies to treat cancer is attacking it with viruses. Virus can kill tumor cells specifically or act as carriers that deliver normal genes into cancer cells. A model for virus therapy of cancer is investigated and some of its predictions are in agreement with results obtained from experimental tumors. Furthermore, the model reveals an oscillatory (periodic or aperiodic) response of tumor cells and virus populations which may difficult clinical prognosis. These results suggest the need for new \textit{in vivo} and \textit{in vitro} experiments aiming to detect this oscillatory response.
\end{abstract}


\maketitle

In the last few years, in which most of the research in molecular biology of cancer focused in new target genes and potential biomarkers, promising roads for tumor treatment such as gene therapy, virus and antiangiogenic drugs have been openned~\cite{Evan,Liotta}. However, most of the clinically-used cancer therapies have been developed empirically~\cite{Evan}, and therefore mathematical models might be complementary (maybe necessary) tools for the understanding of the dynamics of drug response in the organism. Noticeable examples of such modeling are the recent works of Scalerandi \textit{et al.}~\cite{Scalerandi}, in which the action of different antiangiogenic drugs was simulated, of Wu \textit{et al.}~\cite{Wu} and Wein \textit{et al.}~\cite{Wein}, concerning the spreading of a virus that replicates selectively in tumor cells, and of Kozusko \textit{et al.}, addressing cancer cell growth and response to treatment with cyclic-specific chemotherapeutic drugs~\cite{Kozusko}.

Current cancer therapy is based on damaging DNA by irradiation or chemicals. However, cells lacking p53 tumor supressor gene are unable to respond to DNA damage by inducing cell cycle arrest or apoptosis, frequently becoming refractary to chemotherapy or radiation. These cells occur in more than $50\%$ of all human cancers, and hence it is imperative to develope new forms of treatment. Viruses can be used to trigger a DNA damage response without damaging cellular DNA and to selectively replicate in and lyse p53-deficient human tumor cells~\cite{Raj,Bischoff}. Other potential oncolytic agents  were recently tested. Among them are the reovirus, which can be used in more than $50\%$ of all human tumors having an activated Ras signaling pathway~\cite{Coffey}, and the avian Newcastle disease virus~\cite{Lorence}, a genetically altered herpes simplex virus, which kill cancer cells exhibiting an altered p16/pRB tumor suppressor pathway~\cite{Chase}. In particular, clinical trials in humans using Delta-24 RGD adenovirus, an agent with unprecedented success in the treatment of gliomas~\cite{Suzuki}, are currently being designed. Also, since genetically engineered virus can be selected to kill only cancer cells, virus therapy might be more specific than standard DNA-damaging drugs associated to many toxic effects on the normal cells~\cite{Vogelstein}.

In this letter, a model for virus therapy of cancer is investigated and a few of their predictions are compared with results from experimental tumors. The model, based on a previous one recently proposed by us for the avascular growth of tumors~\cite{Silvio} and applied to chemotherapy modelling~\cite{Silvio2}, combines macroscopic reaction-diffusion equations for the nutrients and virus with microscopic stochastic rules governing the actions of individual cancer cells. It seems to be appropriated to study the response of cancers accessible to direct intratumoral virus injection (primary brain tumors and cancers of the head and neck, for example), but mainly their distant metastatic outgrowths. Its prime advantages are the use of both, an individual-based framework in which the fate of each cancer cell is traced, and a set of stochastic rules for cell actions as a simple effective kinetic cellular model describing the population dynamics of interacting cells~\cite{Bellomo,Lachowicz}.

The tissue, represented by a square lattice, is fed by a single capillary vessel at $i=0$, the top of the lattice. The model considers four different cell populations: normal $\sigma_n$, uninfected $\sigma_c$, infected $\sigma_v$ and dead $\sigma_d$ tumor cells. In contrast to the normal and dead tumor cells, one or more cancer cells can pile up in a given site. Thus, the cell populations in any site, with coordinates $\vec{x}=(i,j)$, $i,j=0,1,\ldots,L$, assume one of the possible values $\sigma_n(\vec{x}),\sigma_d(\vec{x})=0,1$ and $\sigma_c(\vec{x}),\sigma_v(\vec{x})=0,1,\ldots$. Initially, a single cancer cell at $\vec{x}=(d,L/2)$ is introduced in the normal tissue. The row $i=L+1$ constitutes the external border of the tissue. The evolution rules implemented in our model are the following:

\textit{Nutrient evolution}.---The nutrients are divided into two classes: directly used, $N(\vec{x},t)$, or not, $M(\vec{x},t)$, for DNA synthesis. Both, $N$ and $M$, are the stationary solutions of the simplest (linear with constant coefficients) dimensionless reaction-diffusion equation
\begin{equation}
\frac{\partial H}{\partial t}=\nabla^2 H-\alpha^2 H \sigma_n -\lambda_H\alpha^2 H (\sigma_c +\sigma_v)
\end{equation}
in which $H$ represents the nutrient fields. Distinct nutrient consumption rates for normal and cancer cells, characterized by factors $\lambda_N$ and $\lambda_M$ ($\lambda_N>\lambda_M$), are assumed. (see ref.~\cite{Silvio} for the complete variable transformations). Also, it is assumed that infected cancer cells sustain their metabolism until lysis.

These equations satisfy periodic and Neumann boundary conditions along the horizontal axis and at the border of the tissue ($i=L$), respectively. At the capillary vessel ($i=0$) $N=M=1$, representing the continuous and fixed supply of nutrients provided by the blood flow.

\textit{Cell dynamics}.---Each cancer cell, randomly selected with equal probability, can carry out one of three actions.

1.\textit{Division}. Uninfected cancer cells divide by mitosis with probability
\begin{equation}
P_{div}(\vec{x})=1-\exp \left[- \left( \frac{N(\vec{x})}{\sigma_c(\vec{x}) \theta_{div}} \right)^2 \right]
\end{equation}
determined by the concentration per cancer cell of the nutrients $N$ necessary for DNA synthesis. The  parameter $\theta_{div}$ controls the shape of $P_{div}$. If the chosen cell is inside the tumor, its daughter will pile up at that site, and $\sigma_c(\vec{x}) \rightarrow \sigma_c(\vec{x})+1$. Otherwise, if the selected cell is on the tumor border, its daughter will occupy at random one of their normal or necrotic nearest neighbor sites $\vec{x^\prime}$. Thus, $\sigma_c(\vec{x^\prime})=1$ and $\sigma_{n,d}(\vec{x^\prime})=0$. In contrast, an infected cancer cell does not divide since its slaved cellular machinery is focused on virus replication.

2.\textit{Migration}. Only uninfected cancer cells migrate with probability
\begin{equation}
P_{mov}(\vec{x})=1-\exp \left[-\sigma_c(\vec{x}) \left( \frac{N(\vec{x})}{\theta_{mov}} \right)^2 \right].
\end{equation}
A selected cell will move to a randomly chosen nearest neighbor site $\vec{x^\prime}$. In terms of cell populations $\sigma_c(\vec{x^\prime}) \rightarrow \sigma_c(\vec{x^\prime})+1$, $\sigma_c(\vec{x}) \rightarrow \sigma_c(\vec{x})-1$, $\sigma_{n,d}(\vec{x})=0$, or $\sigma_{n,d}(\vec{x})=1$ if $\sigma_c(\vec{x})=1$.

3.\textit{Death}. Uninfected cancer cells die with probability
\begin{equation}
P_{del}(\vec{x})=\exp \left[- \left( \frac{M(\vec{x})}{\sigma_c(\vec{x}) \theta_{del}} \right)^2 \right]
\end{equation}
determined by the concentration per cancer cell of the nutrients $M$ present on the cell's microenvironment. This Gaussian distribution has a variance fixed by the parameter $\theta_{del}$. Clearly, if the chosen cell dies $\sigma_c(\vec{x}) \rightarrow \sigma_c(\vec{x})-1$ and $\sigma_d(\vec{x^\prime})=1$ when $(\sigma_c+\sigma_v)$ vanishes.

Infected cancer cells mainly die by lysis, and the death due to insufficient nutrient supply is neglected. So, an infected cell die with probability
\begin{equation}
P_{lysis}(\vec{x})=1-\exp \left( -\frac{T_{inf}}{T_l}  \right)
\end{equation}
where $T_{inf}$ is the time since the infection and $T_l$ is the characteristic period for cell lysis. Again, after lysis $\sigma_v(\vec{x}) \rightarrow \sigma_v(\vec{x})-1$ and $\sigma_d(\vec{x^\prime})=1$ when $(\sigma_c+\sigma_v)$ vanishes.

\textit{Virus spreading}.---The lysis of each infected cancer cell releases $v_0$ virus to the extra-cellular medium. A random fraction of these virus remains on the site of the lysed cell, and the rest is equally distributed among its nearest neighbors. Taken these sources into account, the time evolution of virus concentration is given by the quasi-stationary solutions of the discrete reaction-diffusion equation
\begin{equation}
v(\vec{x},t+1)=v(\vec{x},t)+ \frac{D_v}{4} \sum_{<\vec{x},\vec{x^\prime}>} [v(\vec{x^\prime},t)-v(\vec{x},t)] -\gamma_v v(\vec{x},t).
\label{difvirus}
\end{equation}
Here, $D_v$ is the virus diffusion constant, $\gamma_v$ is the viral clearance rate and the sum, representing the discrete Laplacian on the square lattice, extends over all the nearest neighbors of the site $\vec{x}$. In all simulations, Eq.(~\ref{difvirus}) was iterated $100$ times at each time step, and $D_v$ and $\gamma_v$ were varied on the interval $[0,1]$. Periodic boundary conditions along the y-axis and Neumann boundary conditions at the border of the tissue ($i=L$) and at the capillary vessel ($i=0$) were used for the virus.

Free virus can infect a tumor cell with probability
\begin{equation}
P_{inf}(\vec{x})=1-\exp \left[- \left( \frac{v(\vec{x})}{\sigma_c(\vec{x}) \theta_{inf}} \right)^2 \right],
\end{equation}
a sigmoidal function of the local virus concentration per cancer cell controlled by the parameter $\theta_{inf}$, the efficacy of virus infection.

The virus therapy begins when the tumor attains a growth $N_0$ and consists in a single virus injection. In the direct intratumoral (intravenous) administration an uniforme virus concentration $V=v_0=1$ over the entire tumor (tissue) is supplied. This approach corresponds to the experimental protocols used in severe combined immune deficient (SCID) mice~\cite{Coffey} and \textit{in vitro} essays~\cite{Raj,Bischoff}.

\textit{Results and discussion}.---Simulations in lattices with linear size $L=500$ have been done for each morphology (compact, papillary and ramified) reproduced by our previous model~\cite{Silvio} and commonly observed in real tumors. The corresponding fixed parameters are reported in table I, whereas those controlling virus therapy, namely, $D_v$ (virus diffusion coefficient), $\gamma_v$ (viral clearance rate), $\theta_{inf}$ (efficacy of virus infection) and $T_l$ (characteristic time for cell lysis) were varied. Since the main results are qualitatively the same for all morphology types, only those obtained for compact tumors will be reported here for brevity.

\begin{table}
\caption{Parameter values used in the simulations.}
\begin{tabular}{ccccccl} \hline
Morphology & $\lambda_N$ & $\lambda_M$ & $\alpha$ & $\theta_{div}$ & $\theta_{del}$ & $\theta_{mov}$ \\ \hline
compact    & $25$  & $10$ & $2/L$ & $0.3$ & $0.03$ & $\infty$ (no migration) \\
papillary  & $200$ & $10$ & $2/L$ & $0.3$ & $0.03$ & $\infty$  \\
ramified   & $200$ & $10$ & $3/L$ & $0.3$ & $0.01$ & $\infty$   \\ \hline
\end{tabular}
\end{table}

\begin{figure}
\begin{center}
\resizebox{8.5cm}{!}{\includegraphics{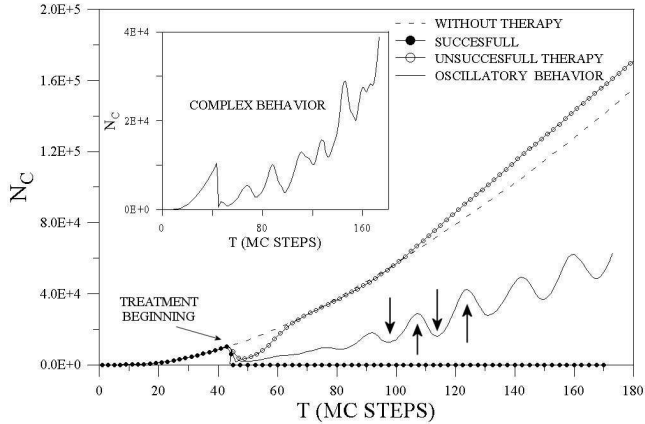}}
\end{center}
\vspace{-0.5cm} \caption{Temporal evolution of the number of cancer cells $N_c$ with and without virus therapy for the compact case.}
\end{figure}

Figure 1 shows the total number of uninfected cancer cells as a function of time for different virus diffusion and clearance rates. Without treatment, the progress in time of cancer cell populations for all the simulated patterns grows accordingly Gompertz curves~\cite{Silvio}. In the case of a succesfull therapy the tumor mass is completely eradicated a few time steps after virus injection. However, it is worthwhile to notice that in addition to the virus oncolytic activity ($\theta_{inf}$ and $T_l$), its spreading properties ($D_v$ and $\gamma_v$) are key features which determine success of the treatment. Indeed, a smaller virus diffusion constant and a greater clearence rate lead to an unsuccesfull therapy. Moreover, after a significant inicial decrease in uninfected cancer cells following virus injection, the tumor grows faster than without treatment. Such result agrees with the claim that cancer cells facing severe threats can trigger adaptation mechanisms as an integrated defense program similar to those observed in bacterial colonies~\cite{Israel}. A tumor submitted to therapeutic approaches which do not lead to its complete eradication may pogressively become more resistant, aggressive and malignant (more invasive), as it is well known in the clinical practice. These results address the central issue concerning the effect of the host immune response on virus therapy, which for human adenoviruses can not be investigated through any appropriate animal model~\cite{Bischoff}. An immune response directed at free virus and late viral antigens expressed on the surface of infected cancer cells can be modeled by an increase in $\gamma_v$ and the introduction of an additional probability to the death of infected cells, respectively. For very small $\gamma_v$ values the tumor is eradicated with an unique dose, in agreement with the essays performed for immune deficient (SCDI)~\cite{Coffey} and athymic~\cite{Bischoff} mice. In turn, for the same dose but higher values of $\gamma_v$, more frequent virus injections are necessary to eradicate the tumor, again as it was observed for syngeneic immune-competent C3H mice in Ref.~\cite{Coffey}. 

\begin{figure}[hbt]
\begin{center}
\resizebox{7cm}{!}{\includegraphics{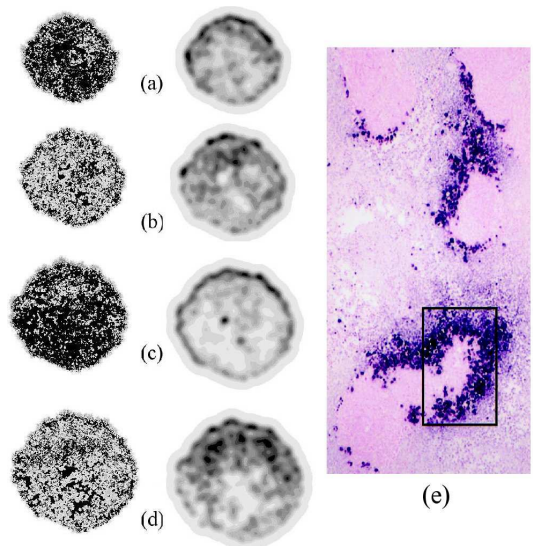}}
\end{center}
\vspace{-0.5cm} \caption{Snapshots of cancer cell (left) and virus (right) patterns at the time steps indicated by the arrows in Fig. 1. Virus and cancer cells concentrations are depicted in gray scale with darker levels corresponding to higher values. Dead cancer cells are depicted in black. In (e) groups of dl1520-infected HCT116 tumor cells (dark cells) bordering uninfected and necrotic tissue are shown. In the boxed area the necrotic cells correspond to the lighter region. (Taken from Ref.~\cite{Heise})}
\end{figure}

A striking and novel result shown in Figure 1 is the possibility of an oscillatory behavior for both virus and cancer cell populations. It contrasts to the long-term, spatially uniform profiles for cells and free virus reached through damped oscillations obtained by Wu \textit{et al.}~\cite{Wu}. Instead, the total number of virus, uninfected and infected cancer cells undergo growing (unstable) oscillations, as shown in Figure 2. Similar patterns exhibiting necrotic regions behind expanding fronts of infected cancer cells have been found by Heise \textit{et al.}~\cite{Heise} (Fig. 2(e)). An oscillatory dynamics may turn innappropriate to monitor the therapy at predefined time intervals, difficulting clinical prognosis. A central feature of such oscillations is that they are not generated by whatsoever time delay between virus infection and cell lysis, as required in Ref.~\cite{Wein}. In fact, our simulations demonstrate that for a vanishing $T_l$ this oscillatory behavior is sustained and, mainly, the effectiveness of virus therapy is largely increased. Furthermore, in this region of parameter space, a single intratumoral virus injection can result in a complete tumor regression with a continously decreasing probability as the clearance rate increases. Thus, the experimentally observed outcomes for reovirus~\cite{Coffey} and adenovirus~\cite{Bischoff} therapies, in which the complete regression of tumors occurs in 60 to 80$\%$ of the mice, seems to correspond to that region of the model parameter space. The complete parameter subspace $\theta_{inf}, \gamma_v$ will be reported elsewhere.

Additionally, our results suggest that virus treatment may be more effective against highly invasive papillary tumors (e. g. glioblastoma multiforme or trichoblastoma), since the cells in such tumors grow slowly due to nutrient limitation. This finding contrast with the conjecture that very low cell densities at the periphery of the ramified tumors may prevent further spreading of the wave of virus infection \cite{Wein}. The differences between the predictions of our model and those in Refs. \cite{Wu,Wein}, namely, oscillations without time delay and sustained wave infection at low cell densities, probably arise from the combined use of macroscopic equations and stochastic microscopic dynamics for discrete tumor cells, instead of partial differential equations (PDEs). The interplay between interactions and fluctuations inherent to the discrete microscopic components of the system may lead to collective behaviors unrevealed by a PDE based model \cite{Louzoun}.

Figure 3 provides further comparison among model predictions and data for virus therapy of tumors in mice. It shows the evolution in time of the simulated number of uninfected tumor cells after virus infection for distinct virus oncolytic activity. The insets show the tumor population and volume size ratios. These ratios are defined as the uninfected cancer cell numbers and tumor volumes normalized by their values at the time of virus injection, respectively. Also, in order to build this figure a time for cancer cells to pass through the DNA synthesis phase of the cell cycle of about $10$ h~\cite{Rew} was assumed, meaning that the time steps in the simulations correspond to about $4-5$ h. In the inset (b) of Fig. 3, the experimental data correspond to the effects of three different virus (a wild-type and UV-inactivated adenovirus, and a mutant adenovirus dl1520) on C33A human cervical tumors xenografts grown subcutaneously in SCID mice~\cite{Bischoff}. These experiments may be simulated in our approach by associating distinct values for $\theta_{inf}$ to different virus, whereas the virus diffusion constants and clearance rates are kept fixed since only one unique experimental model (mouse and implanted tumor) is considered. Remarkably, the predictions of the present model are in good quantitative agreement with the experimental data.

\begin{figure}[hbt]
\begin{center}
\resizebox{8.5cm}{!}{\includegraphics{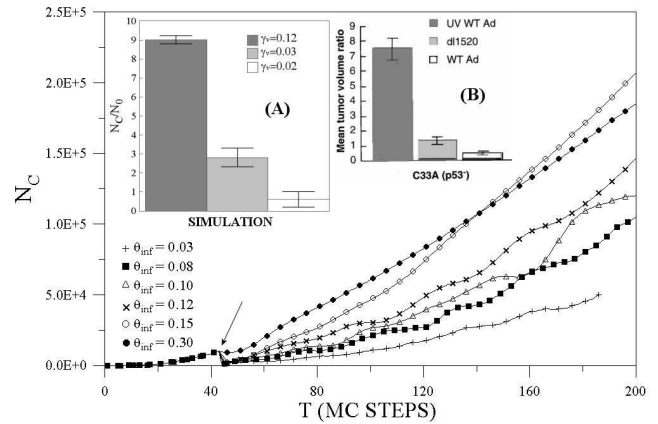}}
\end{center}
\vspace{-0.5cm} \caption{Temporal evolution of the number of uninfected cancer cells for treatments using virus of distinct oncolytic activities. The arrow indicates the time of virus injection. Inset (A) shows the simulated cancer cell population ratios, and the inset (B), taken from Ref.~\cite{Bischoff}, exhibits the C33A cells (p53$^-$) volume ratios after virus injection. In both insets, the ratios 5 weeks after virus injection were  evaluated.}
\end{figure}

In summary, a model combining macroscopic reaction-diffusion equations for the nutrients and virus with microscopic stochastic rules governing the actions of the cancer cells was proposed in order to describe the effects of virus on tumor growth. The simulational results of the model are consistent with the experimental findings for virus therapy. Moreover, they reveal an oscillatory (periodic or aperiodic) response of tumor cells and virus populations on an intermediate region of the model parameter space ($D_v$, $\gamma_v$). In this region, a single intratumoral injection of virus can result in a complete tumor regression with a continously decreasing probability as the clearance rate increases. Such oscillations, which may difficulty clinical prognosis, suggest the need for new \textit{in vivo} and \textit{in vitro} experiments aiming to detect this oscillatory response.

This work was partially supported by the Brazilian Agencies CAPES, CNPq, and FAPEMIG.

\end{document}